\documentclass{emulateapj}

\begin{document}
\slugcomment{Accepted to AJ}

\title{The Century Survey Galactic Halo Project III:  A Complete 4300 deg$^2$ Survey 
of Blue Horizontal Branch Stars in the Metal-Weak Thick Disk and Inner Halo}

\author{Warren R.\ Brown}
\affil{Smithsonian Astrophysical Observatory, 60 Garden St, Cambridge, MA 02138}
\email{wbrown@cfa.harvard.edu}

\author{Timothy C.\ Beers}
\affil{Department of Physics and Astronomy, Center for the Study of
Cosmic Evolution, and Joint Institute for Nuclear Astrophysics,
Michigan State University, E. Lansing, MI  48824}

\author{Ronald Wilhelm}
\affil{Department of Physics, Texas Tech University, Lubbock, TX 79409}

\author{Carlos Allende Prieto}
\affil{McDonald Observatory and Department of Astronomy, 
University of Texas, Austin, TX 78712}

\author{Margaret J.\ Geller,
        Scott J.\ Kenyon,
        \and
        Michael J.\ Kurtz}
\affil{Smithsonian Astrophysical Observatory, 60 Garden St, Cambridge, MA 02138}

\shorttitle{Century Survey Galactic Halo Project III}
\shortauthors{Brown et al.}

\begin{abstract}

	We present a complete spectroscopic survey of 2414 2MASS-selected blue
horizontal branch (BHB) candidates selected over 4300 deg$^2$ of the sky. We
identify 655 BHB stars in this non-kinematically selected sample. We calculate the
luminosity function of field BHB stars and find evidence for very few hot BHB stars
in the field. The BHB stars located at a distance from the Galactic plane $|Z|<4$
kpc trace what is clearly a metal-weak thick disk population, with a mean
metallicity of [Fe/H] $=-1.7$, a rotation velocity gradient of $d
v_{rot}/d|Z|=-28\pm3.4$ km s$^{-1}$ in the region $|Z|<6$ kpc, and a density scale
height of $h_Z=1.26\pm0.1$ kpc. The BHB stars located at $5<|Z|<9$ kpc are a
predominantly inner-halo population, with a mean metallicity of [Fe/H] $=-2.0$ and a
mean Galactic rotation of $-4\pm31$ km s$^{-1}$.  We infer the density of halo and
thick disk BHB stars is $104\pm37$ kpc$^{-3}$ near the Sun, and the relative
normalization of halo to thick-disk BHB stars is $4\pm1$\% near the Sun.

\end{abstract}


 \clearpage
\section{INTRODUCTION}

	Theoretical simulations show that the remnants of hierarchical galaxy
formation in the Milky Way should still be visible as star streams in the stellar
halo \citep{johnston96, harding01, abadi03b, bullock05, font06}. Star counts and
color-magnitude diagrams have proven very effective in identifying structures in the
halo, including the Sagittarius stream wrapping around the sky \citep{majewski03}
and overdensities in Monoceros \citep{newberg02,ibata03,yanny03}, Canis Major
\citep{martin04}, Triangulum-Andromeda \citep{rochapinto04}, Virgo
\citep{duffau06,vivas06,newberg07}, and elsewhere \citep{belokurov06, grillmair06a,
grillmair06b}.  Stellar spectroscopy opens up the possibility of finding structures
in velocity \citep[such as the Sagittarius dwarf galaxy, e.g.][]{ibata94}, in
metallicity, and in distance.  The major difficulty in mapping the stellar halo is
finding tracer stars that are luminous enough to observe at great depths yet common
enough to densely sample the halo.

	In \citet[hereafter Paper I]{brown03}, we introduced the Century Survey
Galactic Halo Project, a photometric and spectroscopic survey from which we selected
Blue Horizontal-Branch (BHB) stars as probes of the Milky Way halo. BHB stars are
evolved, helium core-burning stars that serve as excellent ``standard candles.'' In
\citet[hereafter Paper II]{brown05b}, we explored the Two Micron All Sky Survey
\citep[2MASS]{skrutskie06} and the Sloan Digital Sky Survey \citep[SDSS]{adelman06}
as the basis for a large spectroscopic survey of BHB stars.  In Paper II we
calculated the first field BHB luminosity function, and concluded that field BHB
stars are consistent with populations seen in most globular clusters, but
inconsistent with globular clusters that have substantial extended BHBs.

	Here we describe a complete, non-kinematically selected sample of BHB stars
covering 10\% of the entire sky.  Our survey is inspired by \citet{brown04}, in
which we photometrically selected BHB candidates from the completed 2MASS catalog.
We have now obtained spectroscopy for 2414 2MASS-selected BHB candidates, allowing
us to measure velocities and metallicities for stars to a depth of 8 kpc over a 4300
deg$^2$ region. Such a large-area survey is necessary to unambiguously identify halo
structure: theoretical simulations predict that star streams cover hundreds of
square degrees on the sky \citep{bullock05, font06}.

	Previous spectroscopic surveys of field BHB stars \citep{pier83, sommer89,
arnold92, kinman94, wilhelm99b, kinman04, clewley04, clewley05, kinman07} have
identified BHB stars over relatively small fractions of the sky (10$^2$ - 10$^3$
deg$^2$) compared to the Century Survey Galactic Halo Project.  The exception is the
sample of 1170 BHB stars observed by the SDSS as mis-identified quasars or as filler
objects in low density regions \citep{sirko04a, sirko04b, clewley06}.  In
comparison, our spectroscopic survey of BHB stars is cleanly selected and is 100\%
complete within the selection limits.

	Although our large-area spectroscopic survey is a rich source for general
studies of the thick-disk and inner-halo populations, here we focus our attention on
the properties of the BHB stars.  Our goal is to lay the groundwork for a structure
analysis to be presented in later paper (in preparation).  In \S 2 we describe
spectroscopic observations of stars in the new 4300 deg$^2$ region, and discuss our
selection efficiency for BHB stars.  In \S 3 we discuss the global properties of the
sample, including the mean galactic rotation and metallicity distribution of the
stars.  In \S 4 we calculate the luminosity function of our clean sample of field
BHB stars.  In \S 5 we fit for the density distribution of the BHB stars and
determine the relative normalization of thick disk to inner halo stars.  We conclude
in \S 6.

\section{DATA}

\begin{figure}		
 \includegraphics[width=3.5in]{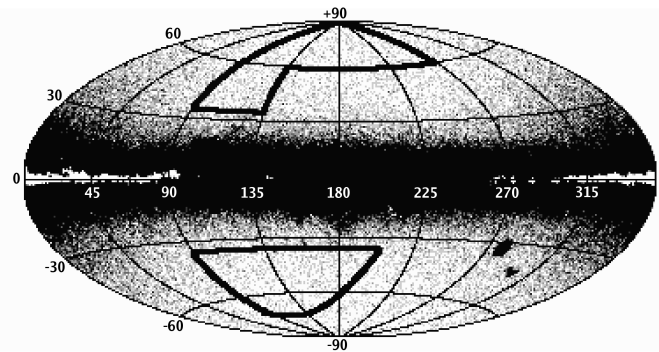}
 \caption{ \label{fig:sky}
	Aitoff sky map in Galactic coordinates, showing the number counts of
2MASS-selected BHB candidates.  Pixels are 1 deg$^2$ in size.  Solid thick
lines indicate our survey regions.  White regions in the disk are regions of
high reddening that are excluded. }
 \end{figure}

\subsection{Target Selection}

	Following \citet{brown04}, we select candidate BHB stars by color from the
2MASS point source catalog \citep{skrutskie06}\footnote{Available at
\url{http://www.ipac.caltech.edu/2mass/}.} with $-0.2 < (J-H)_0 < 0.1$ and $-0.1 <
(H-K)_0 < 0.1$.  We use de-reddened colors and magnitudes \citep{schlegel98} to
create a clean sample.  The color selection is designed to provide a relatively high
selection efficiency ($\sim$40\%) for BHB stars, but a reduced completeness for BHB
stars.  Comparison with the Paper I sample suggests that the color selection samples
$\sim$67\% of the BHB population \citep{brown04}.

	We select BHB candidates in the magnitude range $12.5 < J_0 < 15.5$.  Our
goal is to sample BHB stars as distant as possible, yet at $J=15.5$ the uncertainty
in $(J-H)_0$ exceeds $\pm0.1$ and thus there is little point in going fainter than
$J=15.5$ with 2MASS.  We set $J=12.5$ as our bright limit to avoid thin disk
contamination; a typical BHB star with $M_V=+0.6$ is 2 kpc distant at $J=12.5$.

	Figure \ref{fig:sky}, an Aitoff sky map plotted in Galactic coordinates,
shows our survey region.  The greyscale indicates the number counts of 2MASS BHB
candidates in the magnitude range $12.5 < J_0 < 15.5$.  Our survey region includes
the north Galactic cap opposite the bulge $90\arcdeg<l<270\arcdeg$,
$60\arcdeg<b<90\arcdeg$ plus an extension to $b>35\arcdeg$ at
$90\arcdeg<l<135\arcdeg$.  In the south, our survey samples a similar region bounded
by $b<-35\arcdeg$, $l>90\arcdeg$, and Dec $>-10\arcdeg$.  The survey areas cover
2136 deg$^2$ in the north Galactic hemisphere and 2190 deg$^2$ in the south Galactic
hemisphere.

	There are 2414 BHB candidates in our survey region. The average surface
density of 2MASS-selected BHB candidates is 0.56 deg$^{-2}$.  Our survey is 100\%
complete and can identify stars moving at any radial velocity.

\begin{figure}          
 \includegraphics[width=3.5in]{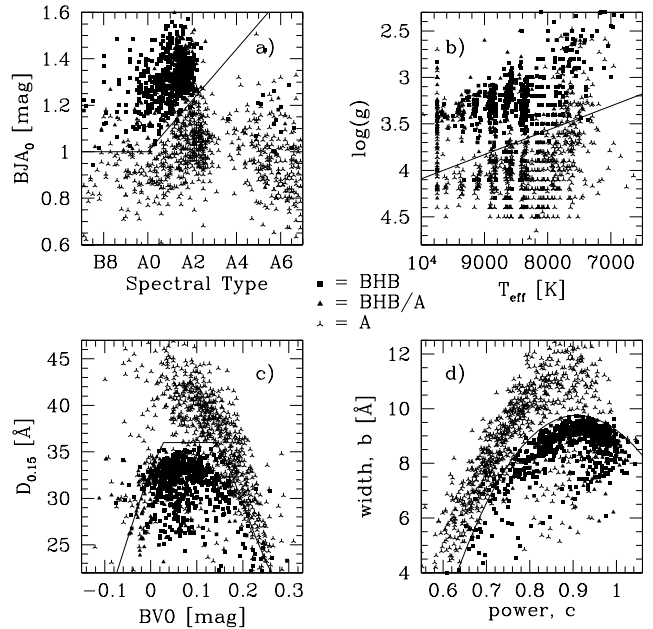}
 \caption{ \label{fig:bhball}
	The four BHB classification techniques applied to our sample:  (a)  
the modified \citet{kinman94} method, (b) the \citet{wilhelm99a} method, (c)  
the \citet{clewley02} $D_{0.15}$-Color method, and (d) the \citet{clewley02}
Scale width-Shape method.  We consider objects satisfying 3 or more of the 4
techniques BHB stars ({\it solid symbols}).}
 \end{figure}

\subsection{Spectroscopic Observations}

	Spectroscopic observations were obtained with the FAST spectrograph
\citep{fabricant98} on the Whipple 1.5m Tillinghast telescope.  Observations were
obtained over the course of 48 nights in 2004 and 2005.  The spectrograph was
operated with a 600 line mm$^{-1}$ grating and a 2 arcsec slit, providing spectral
resolution of 2.3 \AA\ and wavelength coverage from 3450 to 5450 \AA. Exposure
times were chosen to yield a typical signal-to-noise $S/N=30$ in the continuum.

	Paper I contains details of the data reduction.  We use the spectra to
measure radial velocities, spectral types, metallicities, effective temperatures,
and surface gravities of the 2414 BHB candidates.  During the course of this survey
we re-observed 30 objects from Paper I.  The scatter in the spectroscopic
measurements of the same objects provides us with a direct measurement of the
uncertainties:  $\pm16$ km s$^{-1}$ in velocity, $\pm1.2$ sub-types in spectral
classification, $\pm0.4$ dex in [Fe/H], $\pm400$ K in effective temperature, and
$\pm0.3$ dex in surface gravity.

\begin{figure}          
 \centerline{\includegraphics[width=2.25in]{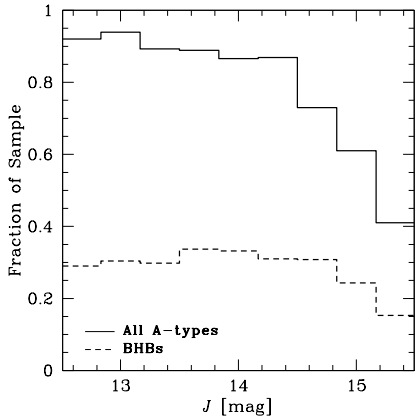}}
 \caption{ \label{fig:frac1}
	Fraction of all A-type stars and BHB stars in our sample as a function
of apparent $J$ magnitude. }
 \end{figure}

\subsection{BHB Classification}

	The major difficulty in using BHB stars as probes of Galactic structure is
the need to distinguish reliably between low surface-gravity BHB stars and higher
surface-gravity A-type dwarfs and blue stragglers.  Although investigators once
thought blue stragglers were a minor component of the halo population, recent
studies \citep{norris91, kinman94, preston94, wilhelm99b, brown03, brown05b}
demonstrate that a surprisingly large fraction of faint stars in the color range
associated with BHB stars are indeed high-gravity stars, many of which are blue
stragglers \citep{preston00, carney05}.

	Our classification of BHB stars is identical to the approach described in
Paper I.  In brief, we apply the techniques of \citet{kinman94}, \citet{wilhelm99a},
and \citet{clewley02, clewley04} to identify low surface-gravity BHB stars.  Figure
\ref{fig:bhball} displays the results for our sample.  We consider objects that
satisfy 3 or more of the 4 classification techniques as BHB stars (solid symbols in
Figure \ref{fig:bhball}); we identify 779 probable BHB stars.

	We expect halo stars to be largely a metal-poor population (e.g., Paper II), yet
124 (16\%) of the BHB stars are relatively metal-rich [Fe/H] $>-0.6$.  Curiously,
the BHB stars with [Fe/H] $>-0.6$ are systematically 0.06 mag bluer in (\bv)$_0$, or
600 K hotter, than the more metal-poor BHB stars.  Hot BHB stars have weak Ca {\sc
ii} K making metallicity measurements difficult.  Furthermore, BHB and main-sequence
A stars have similar surface gravities at $\sim10^4$~K, making classification
difficult.  Thus we consider the [Fe/H] $>-0.6$ BHB stars suspect and mark them as
BHB/A stars in Fig.\ \ref{fig:bhball}.

	Because our goal is to create a clean sample of non thin-disk BHB stars, we
consider the 655 probable BHB stars with [Fe/H] $<-0.6$ as our ``clean'' sample of
BHB stars.  This is consistent with \citet{chiba00}, who use [Fe/H] $<-0.6$ to select
thick-disk and halo stars by metallicity.  We use the clean sample of BHB stars for 
the following analyses.

\begin{figure} 		
 \centerline{\includegraphics[width=2.25in]{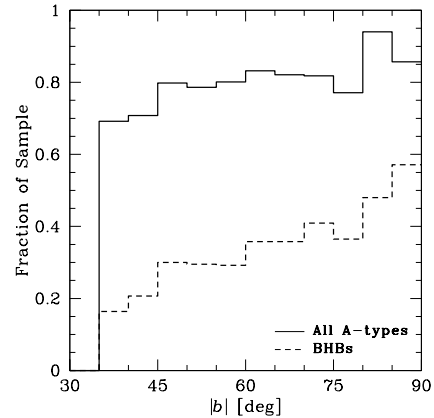}}
 \caption{ \label{fig:frac2}
	Fraction of all A-type stars and BHB stars in our sample as a function
of Galactic latitude $|b|$. }
 \end{figure}

\subsection{Sample Selection Efficiency}

	Our net selection efficiency for BHB stars is 27\% (655 of 2414), and is a
function of both depth and Galactic latitude.  Figure \ref{fig:frac1} plots the
fraction of all stars of spectral type A and the fraction of BHB stars as a function
of apparent $J$ magnitude in our sample.  Our color selection is 90\% efficient for
selecting A-type stars in the interval $12.5 < J < 14.5$, but the efficiency
plummets to 40\% at $J=15.5$ due to increased photometric errors.  Interestingly,
the relative fraction of A-type stars identified as BHB stars increases from 30\% at
$J=13$ to 40\% at $J=15$.  The increasing percentage of BHB stars reflects the
relative fraction of different stellar populations at different depths in the halo.  
This behavior is best illustrated in Figure \ref{fig:frac2}, which displays the
fraction of all A-types and BHB stars in our sample as a function of Galactic
latitude.  BHB stars comprise 20\% of the entire sample with
$35\arcdeg<|b|<45\arcdeg$ and 50\% of the entire sample with
$80\arcdeg<|b|<90\arcdeg$.

	The non A-type stars in our sample are mostly early F-type stars scattering
into our color-selection region, plus a small number of hot subdwarfs and white
dwarfs with blue colors.  We classify 46 objects (2\% of the sample) as subdwarfs
and 21 objects (1\% of the sample) as DA white dwarfs.  One object, CHSS 3842 
(2MASS J010324.54-063210.5), is a hot DB white dwarf.

	We present the photometric and spectroscopic parameters for all 2414 stars in
Appendix A.

\subsection{Variable Stars}

	We don't expect to find RR Lyrae variables in our survey because our color
selection targets stars bluewards of the horizontal branch instability strip.  That
said, our spectroscopy shows that 20\% of the sample is composed of redder, F-type
stars, and our survey is well matched to existing variability surveys such as the
All Sky Automated Survey \citep[ASAS,][]{pojmanski02} and the Northern Sky
Variability Survey \citep[NSVS,][]{wozniak04}.

	We match our entire target list to ASAS and find three variables:  two RR
Lyraes (CHSS 3168 and CHSS 3704) and one eclipsing binary (CHSS 3341).  The NSVS is
better matched to our survey region and survey depth.  We find 1519 stars with NSVS
photometry, of which six have RMS photometric scatter greater than 3 times their
median photometric error.  Of the six possible NSVS variables, two are RR Lyraes
(CHSS 1983 and CHSS 2983) and four show no clear periodicity.  Thus our survey
contains a total of four known RR Lyraes, all of which fall in the reddest 15\% of
the sample.  The four RR Lyraes are not part of the BHB sample.

\begin{figure}          
 \includegraphics[width=3.5in]{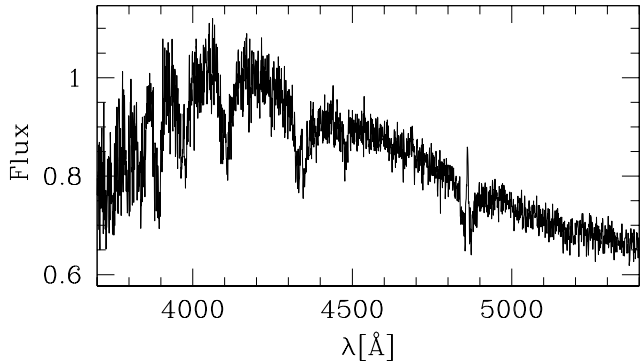}
 \caption{ \label{fig:bestar}
	2MASS J002334.02+065647.6, possibly an accreting white dwarf.}
 \end{figure}

\subsection{A Possible Accreting White Dwarf}


	The most unusual object in the sample is CHSS 3134 (2MASS
J002334.02+065647.6), possibly an accreting white dwarf.  The object is identified
as a DA white dwarf in the \citet{berger80} catalog of blue objects.  Our
higher-resolution spectrum (see Figure \ref{fig:bestar}), however, shows hydrogen
Balmer emission lines in the cores of the absorption lines from H$\beta$ to H10.  
Be stars have similar spectra, but Be stars usually have Balmer emission in only
H$\alpha$ and H$\beta$ \citep[e.g.,][]{bragg02}.  The presence of strong He{\sc i}
4471 indicates the star is hot, with $T_{\rm eff}\gtrsim25,000$ K, and the broad
Balmer absorption lines indicate the star has high surface gravity.  Based on the
observed spectrum (Figure \ref{fig:bestar}), this star is possibly a white dwarf
accreting matter at low rates from a close binary companion.  A white dwarf can
dominate the spectrum if mass transfer has reduced its donor star to almost nothing.  
Alternatively, this star could be a compact binary in which the white dwarf is
illuminating a low-mass companion.  Further spectroscopic follow-up is needed to
establish the exact nature of this unusual system.

\section{GLOBAL PROPERTIES}

\subsection{Disk, Halo Models}

	Given the location and depth of our sample, we expect our BHB stars to
sample both the thick-disk and halo populations.  The halo may not be a single
entity, and the recent \citet{carollo07} analysis of the abundances and kinematics
of $\sim$20,000 stars from SDSS paints a picture of 1) a flattened inner halo with
little or no rotation and with peak metallicity around [Fe/H] $=-1.6$, and 2) a more
spherical outer halo population that is strongly counter-rotating and with a peak
metallicity of around [Fe/H] $=-2.0$.  Our sample of BHB stars reaches heliocentric
distances up to 10 kpc. Thus, in this picture, the majority of our halo stars are
associated with the ``inner-halo'' component of the halo.

\begin{figure}          
 \centerline{\includegraphics[width=2.75in]{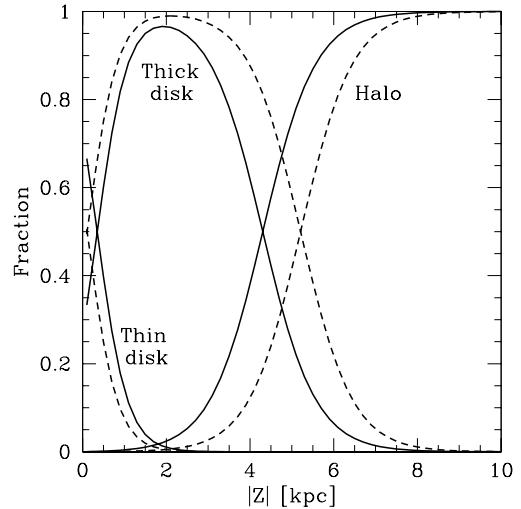}}
 \caption{ \label{fig:profile}
	Relative contribution of thin disk, thick disk, and halo populations for two
representative star count models \citep{siegel02} calculated for our survey volume.  
Solid line:  $Z_{0,thin}=230$ pc, $R_{0,thin}=2$ kpc, $\rho_{thick}=10$\%,
$Z_{0,thick}=600$ pc, $R_{0,thick}=4$ kpc, $\rho_{halo}=0.15$\%, $c/a_{halo}=0.5$,
$r_{halo}^{-3.0}$.  Dashed line:  $Z_{0,thin}=230$ pc, $R_{0,thin}=2$ kpc,
$\rho_{thick}=10$\%, $Z_{0,thick}=600$ pc, $R_{0,thick}=3$ kpc, $\rho_{halo}=1$\%,
$c/a_{halo}=0.7$, $r_{halo}^{-3.5}$.}
 \end{figure}

	To provide context for discussing the properties of our sample, we begin by
investigating the expected contribution of the thick-disk and halo populations.
Figure \ref{fig:profile} shows star-count predictions from \citet{siegel02} for the
relative contribution of the thin disk, the thick disk, and the halo in our survey
volume.  The two sets of lines (solid and dashed) illustrate the range allowed by
their best-fit parameters (see their Table 6).

	Figure \ref{fig:profile} shows that while the relative contribution of the
thick-disk and halo populations in our survey volume is uncertain, the thin-disk
population should be negligible.  The star-count models suggest that the thick disk
should dominate our survey for $|Z|<4$ kpc, while the inner-halo should dominate for
$|Z|>6$ kpc.  Given the uncertainties in the normalizations, however, we simply
conclude that our survey contains a mix of thick-disk and halo stars, and that the
halo contribution increases with distance from the plane.

\subsection{Radial Velocities}

\begin{figure}          
 \includegraphics[width=3.5in]{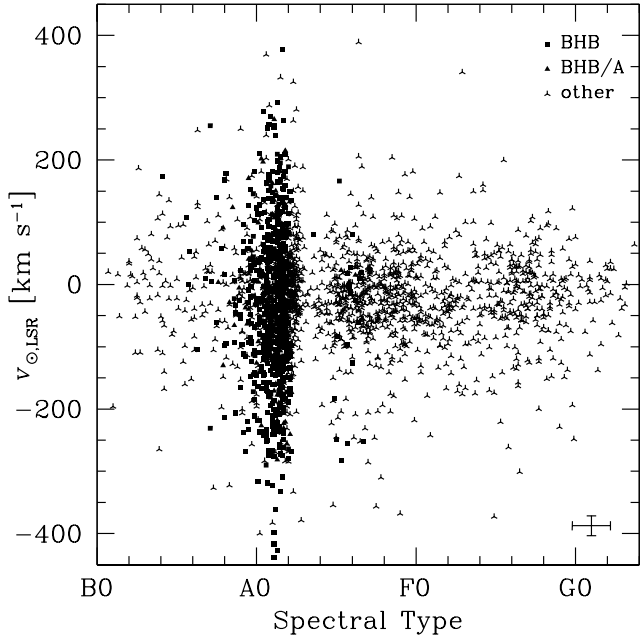}
 \caption{ \label{fig:stvel}
	Spectral types and heliocentric radial velocities with respect to the 
local standard of rest.  Solid symbols mark the BHB stars.  Errorbar indicates
the average uncertainty of the measurements.}
 \end{figure}

	Figure \ref{fig:stvel} displays our heliocentric radial velocities,
corrected for Solar motion relative to the local standard of rest \citep{hogg05}, as
a function of spectral type.  The large group of stars near A0 are the BHB stars
(solid symbols).  The other B-type objects are possibly hot BHB stars, blue
stragglers, or run-away B stars.  The F- and late A-type stars exhibit a smaller
velocity dispersion than the BHB stars, consistent with their being mostly nearby
stars located in the disk.  All of the high velocity stars are probably halo stars.  
Because our sample covers a large area of the sky, we must remove the effects of
Galactic rotation before calculating the line-of-sight velocity dispersion of the
halo stars.  We calculate velocity dispersion and Galactic rotation below.

\subsection{Proper Motions}

	It would be very interesting to know the full space motions of our stars.  
We search existing proper motion catalogs and find 703 matches with the US Naval
Observatory CCD Astrograph Catalog \citep[UCAC2]{zacharias04}, 955 matches with the
US Naval Observatory B1 Catalog \citep[USNOB1]{monet03}, and 2414 matches with the
Guide Star Catalog 2.3 (GSC2.3, B.\ McLean, 2005 private communication).  We proceed
cautiously, however, because our stars are relatively distant, at $2<d<10$ kpc, and
the reported proper motions are typically quite small, $\sim$10 mas yr$^{-1}$.

	We compare proper motions between the UCAC2, USNOB1, and GSC2.3 catalogs.  
The UCAC2 and GSC2.3 proper motions correlate well, but USNOB1 proper motions are
systematically discrepant from the other two catalogs for proper motions less than
10 mas yr$^{-1}$.  The UCAC2 appears the most reliable of the three catalogs
\citep[see][]{mink04}.  However, even the UCAC2 may contain systematic errors on the
scale of degrees (N.\ Zacharias, private communication, 2005), making comparison of
stars in different parts of the sky problematic.  The dispersion of the proper
motions between the three catalogs is $\pm$7 mas yr$^{-1}$; we consider this
estimate a good measure of the accuracy of the catalogs over large areas of sky.

	If we restrict ourselves to proper motions with $>3\sigma$ confidence, the
number of stars with proper motions greater than 20 mas yr$^{-1}$ is approximately
10\% of the catalog matches in all three catalogs.  We estimate tangential
velocities for these stars by combining the proper motions with our distance
estimates (see below).  The formal uncertainty of the tangential velocities is
approximately $\pm300$ km s$^{-1}$.  Because the uncertainty greatly exceeds
expected stellar velocities, these tangential velocity estimates based on the best
proper motions are in practice useless for our analyses.  Thus we ignore proper
motion in the remainder of our paper.

\begin{figure}          
 \includegraphics[width=3.5in]{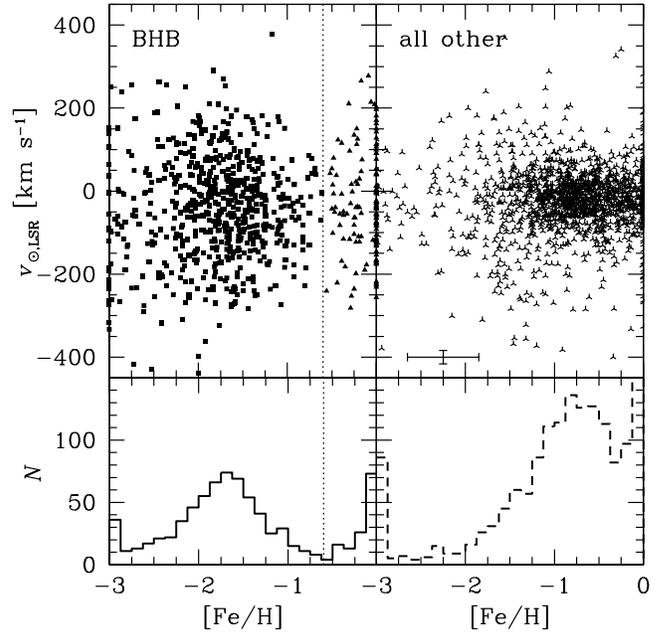}
 \caption{ \label{fig:fehvel}
	Distribution of metallicity [Fe/H] and heliocentric velocity corrected 
to the Local Standard of Rest $v_{\sun,{\rm LSR}}$ for our sample of stars.  
BHB and BHB/A stars are plotted on the left; non-BHB stars are plotted on the 
right.  Errorbar indicates the average uncertainty of the measurements.}
 \end{figure}

\subsection{Metallicities}

	The strongest indicator of metallicity in our A-star spectra is the 3933
\AA\ Ca {\sc ii} K line.  We estimate stellar metallicities based on Ca {\sc ii} as
described in Paper I.  In brief, we use three different techniques:  the spectral
line indices of \citet{beers99}, the equivalent width of Ca {\sc ii} K plus a
chi-square comparison between metallic-line regions in synthetic and observed
spectra \citep{wilhelm99a}, and a Nelder-Mead algorithm that fits the entire
spectrum \citep{nelder65, allende03}.  The final metallicities are the average of
the three techniques and have formal uncertainties of $\pm0.25$ dex.  As mentioned
above, we re-observed 30 objects from Paper I and find that their metallicity 
determinations have a $\pm0.4$ dex RMS scatter.

	Figure \ref{fig:fehvel} plots the metallicities and velocities of the BHB
and BHB/A stars ({\it left panel}) and all the other non-BHB stars ({\it right
panel}) in our sample.  The extra stars at [Fe/H] $=-3$ and 0 are artifacts of our
method; our measurements are restricted to $-3 <$ [Fe/H] $< 0$.

	Examination of Figure \ref{fig:fehvel} reveals that the BHB stars are more
metal-poor than the other stars in our sample.  Excluding the stars at the [Fe/H]
limits, the BHB stars have a median [Fe/H] $=-1.7$.  By comparison, the non-BHB
stars in our sample are more metal-rich, with a median [Fe/H] $=-0.8$.  The
distributions of metallicity and velocity are consistent with the BHB stars
constituting a largely halo population and the non-BHB stars constituting a largely
thick-disk population. 

\subsection{Distances}

	BHB stars are approximate standard candles with luminosities dependent on
effective temperature (color) and on metallicity.  We estimate (\bv)$_0$, which we
label BV0, for our BHB stars using 2MASS photometry and Balmer line strengths (and
SDSS photometry, where available), as described in Paper I.  We then calculate
luminosities for our BHB stars by adapting the $M_V(BHB)$ relation from
\citet{clewley04}.  This relation assumes the {\it Hipparcos}-derived zero point,
$M_V(RR) =0.77\pm0.13$ at [Fe/H] = $-1.60$ \citep{gould98}, a $M_V$-metallicity
slope $0.214\pm0.047$ based on RR Lyrae stars in the Large Magellanic Cloud
\citep{clementini03}, and a cubic relation in (\bv)$_0$ \citep{preston91} to provide
a temperature correction.  A detailed comparison of the luminosity function of field
BHB stars and globular cluster BHB stars in Paper II revealed a systematic 0.3 mag
offset of the Clewley et al.\ $M_V(BHB)$ relation with respect to the globular
clusters.  Therefore, we adjust the zero-point 0.3 mag brighter:
	\begin{equation} \label{eqn:mvbhb}
 M_V(BHB) = 1.252 +0.214{\rm [Fe/H]} -4.423(\bv)_0 +17.74(\bv)_0^2 -35.73(\bv)_0^3.
 \end{equation} Although our zero-point adjustment suggests that the error in the
absolute scale of BHB luminosities may be substantial, we expect that the {\it
relative} BHB luminosities are precise to better than 10\% for our sample.

	There are also 526 stars with early A spectral types between B8 and A3 that
are not BHB stars (see Figure \ref{fig:stvel}).  These high surface-gravity stars
likely comprise a mix of old blue stragglers and young main-sequence stars: two
thirds of the early A-type stars have low mean metallicity [Fe/H] $\simeq-0.9$; one
third are consistent with solar metallicity [Fe/H]=0.  We use the $M_V(A)$ relation
of \citet{sarajedini93} and \citet{kinman94} to estimate luminosities for the 526
early A-type stars and the 124 BHB/A stars:
	\begin{equation} \label{eqn:mvbs}
 M_V(A) = 1.32 + 4.05(\bv)_0 -0.45{\rm [Fe/H]}
 \end{equation} This relation is based on a fit to globular cluster blue stragglers
of similar spectral type.

	We estimate distances using the calculated luminosities and the observed
magnitudes.  We convert 2MASS $J$ magnitudes to $V$ magnitudes by taking our
$(\bv)_0$ estimate and looking up the corresponding $(V-J)_0$ in \citet{kenyon95}
for a star of that color.  This conversion adds additional uncertainty to our
distance estimates.  Thus a typical BHB star with [Fe/H] $=-1.7$ and $(\bv)_0=0.07$
has $M_V(BHB)=+0.65$ and a distance of 9.3 kpc at our limiting magnitude $J=15.5$,
with typical uncertainty of 9\% in distance.  By comparison, a non-BHB, early A-type
star with [Fe/H] $=-0.9$ and $(\bv)_0=0.07$ has $M_V(A)=+2.0$ and a distance of 5.0
kpc at our limiting magnitude, with a typical uncertainty of 12\% in distance.  
These distance uncertainties do not include systematic errors.

	Stars of later spectral type than the BHB / early A stars are intrinsically
less luminous objects in the nearby disk; we exclude these objects from our analysis
and do not calculate their luminosities and distances.

\begin{figure}          
 \includegraphics[width=3.5in]{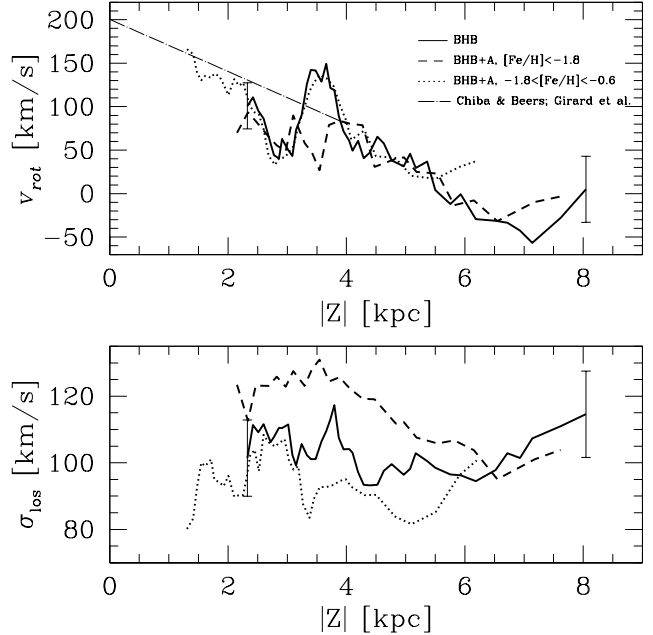}
 \caption{ \label{fig:rot}
	Mean rotation ({\it upper panel}) and line-of-sight velocity dispersion
({\it lower panel}) of 1225 BHB and early A stars located $8<R<11$ kpc ({\it solid
lines}).  We divide the sample into thirds by [Fe/H] and find that the most
metal-rich third [Fe/H] $>-0.69$ ({\it dotted lines}) has systematically higher
rotation velocity and lower velocity dispersion than the most metal-poor third
[Fe/H] $<-1.63$ ({\it dashed lines}).}
 \end{figure}

\subsection{Mean Galactic Rotation}

	Previous surveys provide conflicting measurements of the stellar halo
rotation:  it may be prograde \citep{chiba00, sirko04b}, retrograde
\citep{majewski92, majewski96, carney96, spagna03, kinman04, kinman07, carollo07},
or nothing at all \citep{layden96, gould98, martin98, gilmore02, brown05b}.  
Curiously, measurements of retrograde rotation mostly come from surveys of the north
Galactic pole.  Differences in the observed halo rotation may also arise from the
manner in which different samples have selected from inner-halo and outer-halo
populations, which \citet{carollo07} argue have quite different rotation
characteristics. Our large area survey is ideal for measuring the bulk rotation of
the inner stellar halo.  Our stars are selected photometrically without any
kinematic bias, and our survey is 100\% complete over two large contiguous regions.

	We calculate mean rotation using the formalism of \citet{frenk80}.  We
assume the stars are in pure rotation with uniform velocity about the rotation axis
of the Galactic disk, and project the observed radial velocity onto the azimuthal
(rotation) direction.  The \citet{frenk80} formalism provides an estimate of both
rotation velocity $v_{rot}$ and the line-of-sight dispersion $\sigma_{los}$ for a
set of stars.

	Figure \ref{fig:rot} plots the results of this analysis as a function of
$|Z|$.  We bin in $|Z|$ by binning together 78 stars ordered in $|Z|$, and moving
through the sample in steps of 10 stars.  We adopt this approach to avoid any
effects of arbitrary placement of bins on the results; bins are typically $\sim$0.6
kpc in size, growing to $>1$ kpc at $|Z|>5$ kpc.  We consider three different
samples of stars:  the clean sample of 655 BHB stars (solid line), a combined sample
of 344 metal-poor BHB and early A-type stars with [Fe/H] $<-1.8$ (dashed line), and
a combined sample of 541 intermediate metallicity BHB and early A-type stars with
$-1.8<$ [Fe/H] $<-0.6$ (dotted line).  The latter two metallicity cuts are intended
to select halo and thick disk stellar populations, respectively, following
\citet{chiba00}.

	All stars rotate well below the solar value; $v_{rot}$ drops monotonically
with $|Z|$.  This conclusion is valid for stars in the region $8<R<12$ kpc and
$2<|Z|<9$ kpc. A linear least squares fit to the clean sample of BHB stars
yields $v_{rot(BHB)} = (-28\pm3.4) |Z| + (175\pm16)$ km s$^{-1}$. The observed
velocity gradient is statistically identical to the $-30\pm3$ km s$^{-1}$
gradient found for thick disk stars in the region $0<|Z|<4$ kpc by
\citet{chiba00} and by \citet{girard06} (see Fig.\ \ref{fig:rot}).  
What is remarkable, however, is that all three of our samples show the same
velocity gradient. The metal-poor stars have only a marginally shallower
velocity gradient and lower zero-point $v_{rot(metal~ poor)} = (-20\pm3) |Z| +
(129\pm13)$ km s$^{-1}$ than the intermediate-metallicity stars with
$v_{rot(intermediate~ metallicity)} = (-24\pm4) |Z| + (166\pm13)$ km s$^{-1}$.
Thus the mean rotation velocities suggest that our samples contain significant
numbers of thick disk stars with $|Z|\lesssim5$ kpc.

	The BHB and A stars located at $1.5<|Z|<3$ kpc have $v_{rot}\sim100$ km
s$^{-1}$, consistent with the rotation measured from faint F and G stars at similar
distances \citep{gilmore02,wyse06}.  \citet{gilmore02} argue that a single, coherent
thick disk should have a constant rotation velocity far from the plane, and thus the
observed intermediate $\sim$100 km s$^{-1}$ rotation is evidence for a merger origin
for the thick disk.  However, our sample clearly contains a mix of stellar
populations with different kinematics:  the metal-poor sample has a systematically
larger line-of-sight velocity dispersion than the intermediate metallicity sample at
a given $|Z|$.

	The lower panel of Figure \ref{fig:rot} plots the line-of-sight velocity
dispersion $\sigma_{los}$ of the stars as a function of $|Z|$.  The line-of-sight
velocity dispersion of the clean BHB sample is statistically consistent with a
constant value of $103\pm6$ km s$^{-1}$.  This is consistent with the $\sim$100 km
s$^{-1}$ velocity dispersion measured for thick disk stars at the same depth towards
the south Galactic pole \citep{girard06}.  Yet the metal-poor sample of stars has
$\sigma_{los}=+117\pm10$, consistent with a more halo-dominated population.  
Clearly, our BHB stars are a mix of thick-disk and halo populations.  We note that the
drop in velocity dispersion at $|Z|>4$ kpc seen in the metal-poor stars, though not
statistically significant, may in fact be a real feature due to velocity structure
in the sample.

	We find no evidence for significant rotation of the inner halo in the range
$5.5<|Z|<9$ kpc.  Figure \ref{fig:rot} shows that the formal $v_{rot}$ values dip
below zero in the region $6<|Z|<8$ kpc, but the uncertainties are large (note the
error bars in Fig.\ \ref{fig:rot}).  Interestingly, the final bins of both the clean
BHB sample and the [Fe/H] $<-1.8$ BHB and A star sample have values around zero.  
In the volume $5.5<|Z|<9$ kpc, the clean BHB sample contains 131 stars with a formal
$v_{rot}=-4\pm31$ km s$^{-1}$, while the metal-poor sample contains 76 stars with a
formal $v_{rot}=-3\pm37$ km s$^{-1}$.

\begin{figure}          
 \centerline{\includegraphics[width=3.0in]{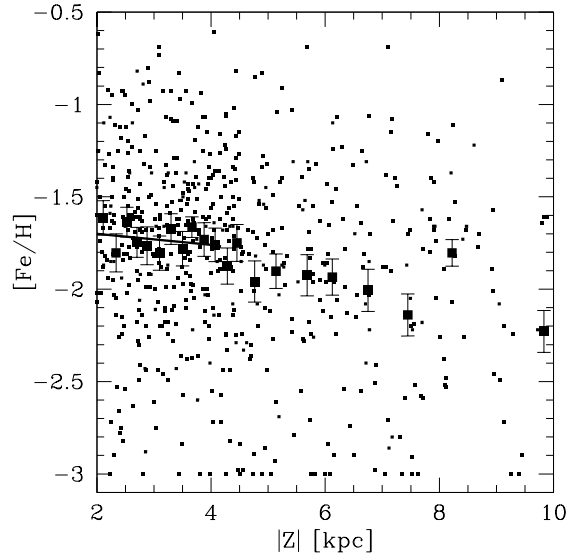}}
 \caption{ \label{fig:fehzz}
	Position and metallicity of our clean sample of BHB stars.  Solid squares
are the average metallicity found in bins of 50 stars.  A linear least squares fit
of the thick disk metallicity gradient $2<|Z|<4$ kpc finds d[Fe/H]/d$Z=-0.03 \pm 
0.05$.  The mean metallicity of BHB stars with $5< |Z| < 9$ kpc is [Fe/H]$=2.0$.}
 \end{figure}

\subsection{The Metal-Weak Thick Disk}

	The thick disk is generally understood to have a metallicity distribution
that peaks around [Fe/H] $\sim-0.7$ \citep[e.g.][]{gilmore95,allende06}.  This
conclusion is at odds with the low metallicity of our BHB stars located at $|Z|<4$.
Figure \ref{fig:fehzz} plots the position $Z$ and metallicity of our clean
sample of BHB stars.  Solid squares are the average metallicity found in bins of 50
stars.  BHB stars located at $2<|Z|<4$ kpc have an average metallicity of
[Fe/H] $=-1.7$; including the BHB/A stars increases the average metallicity only
slightly, to [Fe/H] $=-1.4$.  The average metallicity of BHB stars in our sample
with $5<|Z|<9$ kpc is [Fe/H] $=-2.0$. The small numbers of stars in this region
prevent a determination of a significant metallicity gradient, however it is
interesting that the average metallicity is {\it lower} than one might have
expected from a canonical halo population with peak metallicity [Fe/H] $= -1.6$
\citep{carney96}. It may be that we are seeing evidence for the transition
of inner- to outer-halo populations in this interval, as suggested by 
\citet{carollo07}.

	One explanation for both the low metallicity and the thick-disk-like
kinematics (Figure \ref{fig:rot}) of our BHB stars at $|Z|<4$ kpc is provided by the
existence of a metal-weak thick disk \citep[e.g.,][]{norris85, morrison90, chiba00,
beers02}. BHB stars are associated with metal-poor populations, such as globular
clusters, thus it may be that our survey for BHB stars preferentially traces the
metal-weak thick disk.  In any case, the existence of very metal-poor BHB stars with
thick-disk kinematics presents another clue for formation scenarios of the Milky
Way.

	A linear least-squares fit to the BHB stars located at $2<|Z|<4$ kpc reveals
a weak metallicity gradient, d[Fe/H]/d$Z=-0.03 \pm 0.05$ (solid line in Figure
\ref{fig:fehzz}), consistent with zero.  Previous studies find no evidence for a
vertical metallicity gradient in the more commonly studied (metal-rich) thick disk
\citep{gilmore95, allende06}.

\section{LUMINOSITY FUNCTION OF BHB STARS}

	Understanding the luminosity function of field BHB stars is important for
interpreting our maps of the Galactic halo.  BHB stars have a distribution of
intrinsic luminosities, thus we sample different luminosity BHB stars to different
depths.  The luminosity function describes the number of BHB stars per unit volume
in the luminosity interval $M_V$ to $M_V+dM$.  While we must infer a star's
luminosity from its color and metallicity, the $M_V(BHB)$ relation (Eqn.\
\ref{eqn:mvbhb}) specifies only how a particular color and metallicity map to a
particular $M_V$.  We emphasize that it is the {\it observed distribution} of BHB
colors and metallicities that determines the form of the luminosity function.  
Paper II presents a more extensive discussion of this issue and the underlying
physics.

	We calculate the field BHB luminosity function using the \citet{eep}
non-parametric maximum-likelihood method.  An important feature of this method is
that the density terms drop out, thus the luminosity function calculation is
unbiased by density variations.  In other words, the maximum-likelihood method does
not require knowledge of the halo density distribution $\rho(R,Z)$, it only requires
that the luminosity function is independent of position in the volume sampled.  
Because stellar density varies with position in the Milky Way, we compute only the
{\it form} of the luminosity function and arbitrarily normalize the luminosity
function to one.

\begin{figure}          
 \includegraphics[width=3.5in]{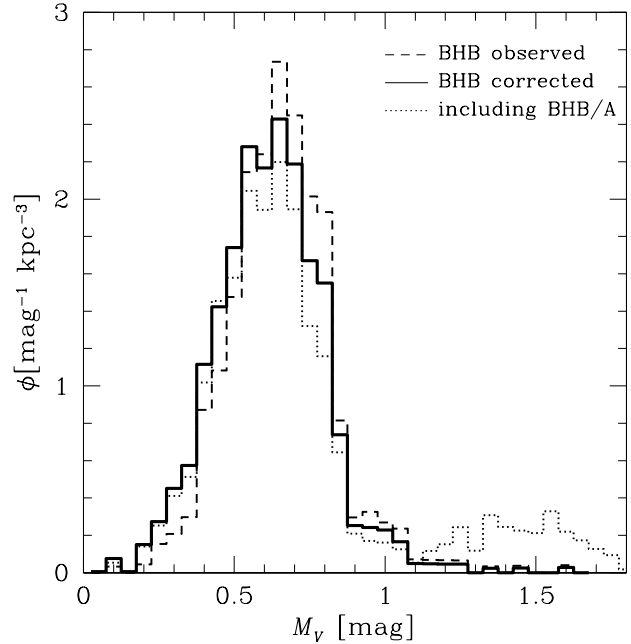}
 \caption{ \label{fig:lf}
	Luminosity function of field BHB stars, calculated for the observed sample
(dashed line) and corrected for completeness (solid line).  If we include the BHB/A
stars with [Fe/H] $>-0.6$ (dotted line) a faint tail appears.  However, it is clear
that hot, intrinsically faint extended BHB stars are not a significant fraction of
the field BHB population.}
 \end{figure}

	Figure \ref{fig:lf} plots the luminosity function of the clean sample of BHB
stars (dashed line).  The luminosity function rises steeply at bright luminosities,
peaks at $M_V=0.64$, and falls rapidly with a tail at faint luminosities.  Although
our statistics are greatly improved over the field BHB luminosity function measured
in Paper II, we caution that our BHB sample is incomplete for stars $(J-H)_0>0.1$.

	In principle, we can correct for our sample incompleteness.  The Paper I
sample is complete over a much broader range of color than our 2MASS-selected
sample.  Thus the Paper I sample can provide us with the distribution of BHB colors
with $(J-H)_0>0.1$ that are missing from our 2MASS-selected sample.  We estimate the
luminosities of the missing, redder BHB stars as follows.  First, we determine the
distribution of BV0 colors of our stars as a function of $(J-H)_0$.  Second, we
determine the distribution of [Fe/H] for BHB stars with colors near $(J-H)_0=0.1$.  
Third, we construct cumulative distributions of BV0 and [Fe/H] from our
observations, and then sample these distributions to obtain the expected
distribution of $M_V$'s for the missing stars.  Finally, we correct the luminosity
function bins for the appropriate fraction of missing stars as determined from the
Paper I sample.

	Figure \ref{fig:lf} plots the BHB luminosity function corrected for
incompleteness (solid line).  The effect of the incompleteness correction is to
increase the fraction of redder, more luminous stars, and thus shift the peak of the
distribution to $M_V=0.60$.  We also plot the corrected luminosity function for the
combined sample of BHB and BHB/A stars.  Because the BHB/A stars have [Fe/H] $>-0.6$
and bluer colors, on average, than the BHB stars, they are intrinsically
under-luminous and fall entirely in the faint tail of the luminosity function.  
These stars are possibly hot extended BHB stars, though such a strong preference for
high metallicities is not observed in globular clusters.  For example, NGC 7078 has
[Fe/H] $=-2.25$ \citep{harris96} and contains a large number of extended BHB stars.  
Even if the BHB/A stars are all BHB stars, it is clear from Figure \ref{fig:lf} that
extended BHB stars are not a significant fraction of the field BHB population.

\begin{figure*}          
 
\centerline{\includegraphics[width=2.5in]{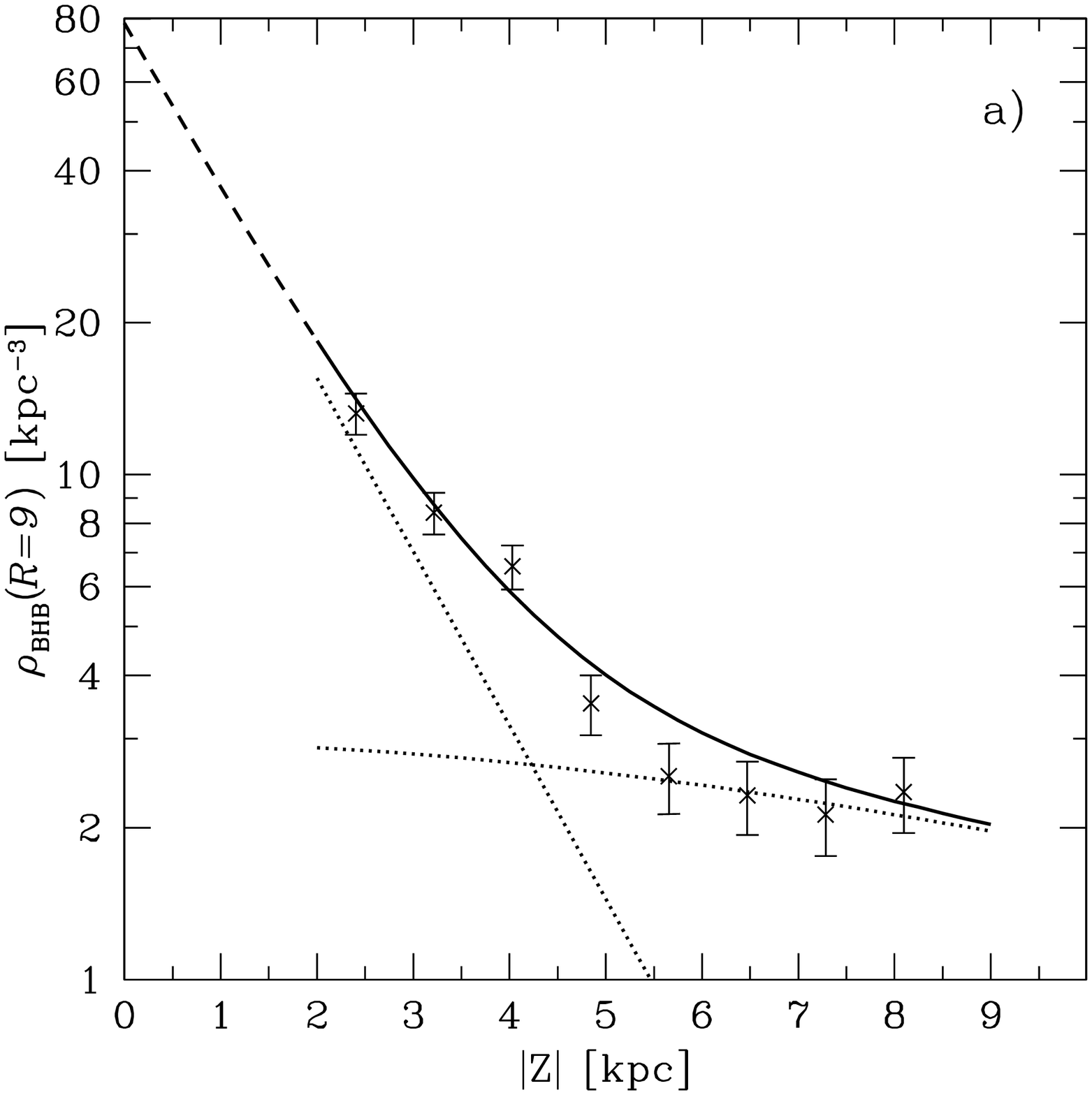}~~~~~~~~~~~~~~~~~~~~~~\includegraphics[width=2.5in]{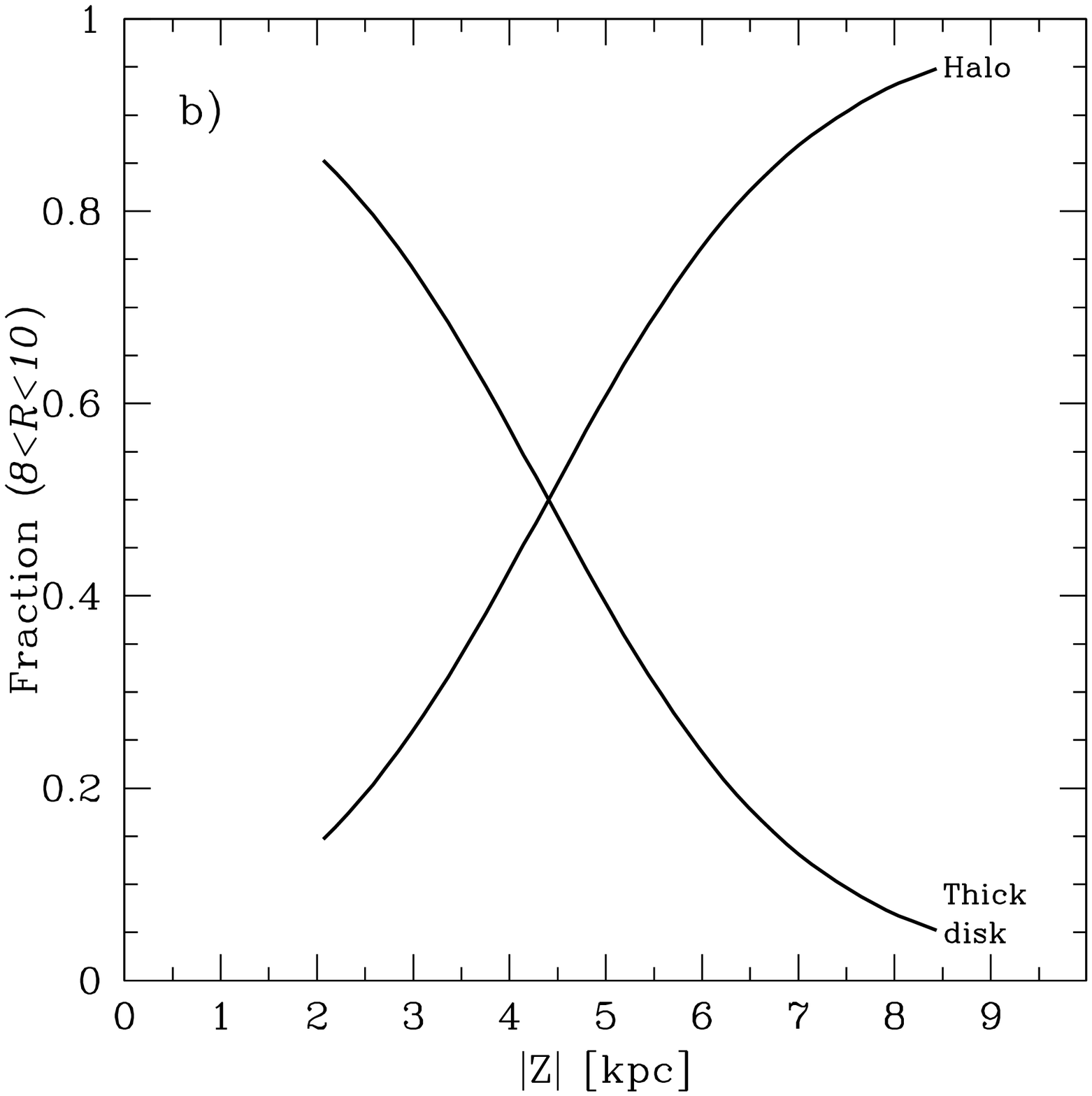}}
 \caption{ \label{fig:modfinal}
	a) Best-fit density distribution (solid line) and the observed density of
BHB stars at $R=9$ kpc (points).  The extrapolation (dashed line) suggests there are
$78\pm30$ kpc$^{-3}$ halo and thick disk BHB stars at $(R, Z)=(9, 0)$ kpc.  b) The
fraction of thick disk and halo BHB stars in our survey volume located $8<R<10$ kpc.}
 \end{figure*}

\section{DENSITY DISTRIBUTION OF FIELD BHB STARS}

	Our sample of BHB stars can potentially provide an excellent measure of the
density distribution of the thick-disk and inner-halo stellar populations.
Traditionally, the Galaxy's density distribution is measured with star counts. The
star-count technique is powerful because can use photometric catalogs containing
millions of stars \citep[e.g.][]{siegel02, larsen03}. However, stellar populations
are a complex function of both color and apparent magnitude. Thus star-count
techniques suffer from uncertainties in stellar color-luminosity relations, as well
as contamination from binaries and non-stellar objects. By comparison, our survey
provides a very clean sample of spectroscopically identified BHB stars. Although the
numbers of BHB stars is much smaller than samples of stars used by star counts, our
spectra provide precise metallicity and distance determinations for every star.

	We begin by considering the volume of space sampled by our survey.  Using
the BHB luminosity function, we calculate the fraction of stars at a given distance
that fall within our survey magnitude limits.  We expect that our survey is more
than 50\% complete for BHB stars in the range $2.5 < d < 9.5$ kpc.  This range of
heliocentric distance samples the region $8<R<12$ kpc and $2<|Z|<8.5$ kpc for our
predominantly high Galactic-latitude survey region.  Thus we restrict our density
distribution analysis to the above ranges of $d$, $R$, and $Z$.  There are 544 BHB
stars that fall within these ranges.

	Before calculating the density distribution, we correct the observed BHB
sample for incompleteness.  First, the BHB luminosity function tells us the fraction
of stars missing at each distance $d$.  Because of our restriction in distance, this
correction applies to only a handful of stars near the boundaries of the sample.  
Second, the ratio of the observed and corrected luminosity functions tells
us the fraction of BHB stars missing at each $M_V$ because of our color selection;  
we weight stars at a given $M_V$ appropriately.  Finally, we use Figure
\ref{fig:frac1} to estimate the fraction of faint BHB stars missing because of
increased photometric errors, and give additional weight to stars with $J>14.5$.

	We assume the density distribution is a sum of a thick-disk and halo
population with the canonical forms:
  \begin{equation} \label{eqn:density}
\rho(R,Z) = \rho_{0,thick} \exp{(-Z/h_Z)} \exp{(-R/h_R)} + \rho_{0,halo} / (a_0^n + 
R_g^n)
  \end{equation} where $R_g = \sqrt{R^2 + (z~c/a)^2}$ and $c/a$ is the halo axial
ratio.  These are the same relations used by \citet{siegel02} in their fits to star 
counts.  Selecting a power-law halo instead of a de Vaucouleurs halo is mostly a
cosmetic choice; our BHB sample provides very little leverage on the halo density
profile.  Our sample is a high Galactic-latitude sample best suited to measuring
$h_Z$ and the relative normalizations of thick disk and halo BHB star densities.

	\citet{girard06} caution that distance uncertainties, convolved with the
sharply falling density distribution of stars, can alter the ``observed'' density
distribution from the actual, intrinsic form.  Thus, we mimic their procedure, and
artificially partition each star into 100 positions in distance, with a distribution
of distances described by a Gaussian distribution around the best value for each
star.  We then apply our limits in $d$, $R$, and $Z$ to the subunits.  This
procedure allows stars that would otherwise be excluded by distance cuts to
contribute a small amount of weight appropriate to the uncertainty in their distance
estimate.  We bin the subunits into volumes at fixed intervals $R$ and $Z$, and
perform our density fits to these bins using $\chi^2$ minimization techniques
\citep{press92}.

	We start by testing fits to the different components of our BHB sample.  
The density distribution of stars located at $-4<Z<-2$ kpc and $2<Z<4$ kpc are very
similar.  In the final fit we consider stars above and below the plane together as a
function of $|Z|$.  We try different bin sizes and different ranges of $R$ and $Z$,
and find that stars located at $2<|Z|<4$ kpc prefer thick-disk scale lengths in the
ranges $2.5<h_R<4$ kpc and $1<h_Z<2$.  Unfortunately, stars located farther out, at
$5<|Z|<8.5$ kpc, provide very little constraint on the form of the halo profile.  If
we fix the core radius to $a_0=6.3$ kpc \citep{girard06} and the halo axial ratio to
$c/a=0.7$ \citep{robin00,siegel02}, our sample prefers an halo power-law index in
the range $2.5<n<3$.

	We fit Equation \ref{eqn:density} to the full BHB sample, holding $c/a=0.7$
fixed and fitting the other 6 parameters.  Table \ref{tab:fit} gives the best-fit
parameters and Figure \ref{fig:modfinal} shows the results.  Figure
\ref{fig:modfinal}a plots the observed density of BHB stars at $R=9$ kpc, the
best-fit density distribution (solid line), and the thick-disk and halo components
(dotted lines).  Figure \ref{fig:modfinal}b plots the fraction of thick-disk and
halo stars in our survey volume located $8<R<10$; the components contribute equal
fractions of BHB stars at $|Z|\sim4.5$ kpc.

	We map out contours of $\chi^2$ to understand the uncertainty in our
best-fit parameters.  We caution that our sample provides little constraint on
$a_0$, and that there is a significant degeneracy between the thick-disk scale
length, $h_R$, and the halo power law index $n$.  Star-count models give similar
results to our best-fit halo power-law index $n\sim2.5$ \citep[e.g.]{robin00,
siegel02}.  In contrast, \citet{chiba00} find $n\sim3.5$ from their spectroscopic
sample of metal-poor stars.  Given the uncertainties, however, our BHB stars provide
no significant constraint on the shape of the halo.  What we can measure with
certainty is the normalization of halo to thick disk BHB stars.

	The density of halo and thick disk BHB stars is $104\pm37$ kpc$^{-3}$ near
the Sun $(R, Z) = (8, 0)$ kpc, in good agreement with \citet{green93}'s lower limit
of $51\pm17$ kpc$^{-3}$.  \citet{kinman94} report a three times smaller density of
30 kpc$^{-3}$, but their sample has little constraint on thick disk BHB stars that
dominate the BHB density near the Sun.  We find that the relative normalization of
halo to thick disk BHB stars is $4\pm1\%$ near the Sun.

\begin{deluxetable}{lcl} 	
\tablewidth{0pt}
\tablecaption{BHB DENSITY DISTRIBUTION\label{tab:fit}}
\tablehead{ \colhead{Param} & \colhead{Value} & \colhead{Units} }
	\startdata
$\rho_{0,thick}$& $ 960  \pm 170 $  & kpc$^{-3}$ \\
$h_Z$		& $ 1.26 \pm 0.1 $  & kpc \\
$h_R$		& $ 3.5  \pm 0.5 $  & kpc \\
$\rho_{0,halo}$	& $1040  \pm 180 $  & kpc$^{-3}$ \\
$a_0$		& $ 5.8  \pm 3   $  & kpc \\
$n$		& $ -2.5 \pm 0.5 $  & \\
$c/a$		& $\equiv 0.7$      & \\
	\enddata 
 \end{deluxetable}

\begin{figure}          
 \centerline{\includegraphics[width=2.0in]{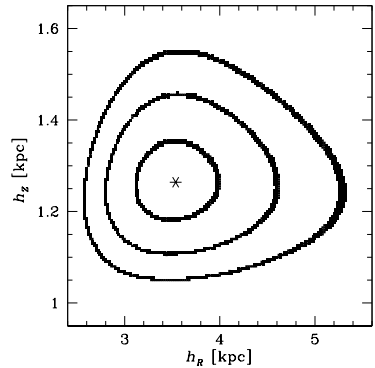}}
 \caption{ \label{fig:chisq}
	$\chi^2$ contours for thick disk scale lengths; the asterisk marks our
best-fit values for $h_Z$ and $h_R$.  We calculate these contours by fixing the the
halo power law and fitting the thick disk and halo normalizations.  The inner
contour matches our best estimate of 1$\sigma$ significance based on boot-strap
resampling. }
 \end{figure}

	Our high-latitude sample also provides a good constraint on the vertical
density distribution of BHB stars in the thick disk. To obtain a self-consistent
picture of the relationship between the thick-disk scale lengths $h_Z$ and $h_R$, we
fix the form of the halo power law (see Table \ref{tab:fit}), vary $h_Z$ and $h_R$
across a grid of values, and fit only the normalizations. This approach results in
contours of $\chi^2$ illustrated in Figure \ref{fig:chisq}. The contours do not
correspond to exact significance levels, but we have chosen the inner contour to
match our best estimate of 1$\sigma$ significance based on boot-strap resampling.
The asterisk in Figure \ref{fig:chisq} marks our best-fit values of $h_Z$ and $h_R$.

	Previous star-count models for the thick disk find either a large scale
height (1.2 - 1.4 kpc) and low normalization \citep{gilmore83, morrison00, reid93,
juric05} or a smaller scale height (0.75 - 1.0 kpc) and high normalization
\citep{robin96, siegel02, robin03, cabrera05, du06, girard06}. Our scale height
$h_Z=1.26\pm0.1$ is consistent with the larger scale heights.

	The thick-disk scale length is interesting because it determines the
rotational-equilibrium of thick disk stars.  Star-count estimates range from
$2.5\pm0.3$ kpc \citep{robin96, robin03} to $4.3\pm07$ kpc \citep{larsen03}.  Our
scale length, $h_r=3.5\pm0.5$ kpc, falls in the middle of this range, similar to
determinations from \citet{siegel02} and \citet{juric05}.  This agreement shows the
power of a clean, even if small, spectroscopic sample.  If our BHB stars preferentially
trace the metal-weak thick disk, then the density parameters in Table 1 reflect the
density distribution of the metal-weak thick disk.

\section{CONCLUSIONS}

	We discuss a complete spectroscopic survey of 2414 2MASS-selected BHB
candidates over 4300 deg$^2$ of sky.  We identify 655 BHB stars in this
non-kinematically selected sample.  The luminosity function of the field BHB stars
has a median value of $M_V=0.65$ and a small tail extending to $M_V>1$, but shows
very few extended BHB stars.

	The BHB stars located at $|Z|<4$ kpc have a mean Galactic rotation and
density distribution remarkably consistent with a metal-weak thick-disk population.  
The $|Z|<4$ kpc BHB stars have a mean metallicity of [Fe/H] $=-1.7$, a velocity
gradient of $d v_{rot}/d|Z|=-28\pm3.4$ km s$^{-1}$, and a vertical scale height of
$h_Z=1.26\pm0.1$ kpc.  We infer a space density of $100\pm36$ kpc$^{-3}$ thick disk
BHB stars near the Sun.  RR Lyrae stars, by comparison, have a much less prominent
disk component near the Sun \citep{martin98}.  The existence of metal-poor BHB stars
with thick-disk kinematics and scale heights present another clue for formation
scenarios of the Milky Way.

	The BHB stars located at $5<|Z|<9$ kpc have a mean Galactic rotation and
density distribution consistent with a predominantly halo population. The halo BHB
stars have a mean metallicity of [Fe/H] $=-2.0$, a mean Galactic rotation of
$-4\pm31$ km s$^{-1}$, and relative normalization of $4\pm1$\% with respect to the
thick disk near the Sun $(R, Z)=(8, 0)$ kpc.  This is the best currently
available measurement of the relative normalization of the halo and thick disk, an
important quantity that enters into all models of the local structure of the Galaxy.

	In the future, having established the global properties of our survey, we
hope to analyze the BHB sample for structure in space, velocity, and metallicity.

\acknowledgements

	We thank Perry Berlind and Mike Calkins for their dedicated observing at the
Whipple 1.5 m Tillinghast telescope.
	This project makes use of data products from the Two Micron All Sky Survey,
which is a joint project of the University of Massachusetts and the Infrared
Processing and Analysis Center/Caltech, funded by NASA and the NSF.
	This project makes use of NASA's Astrophysics Data System Bibliographic
Services.
	This work was supported in part by W.\ Brown's Clay Fellowship and the 
Smithsonian Institution.
	T.C.B.\ acknowledges partial funding for this work from grants AST 04-06784,
AST 06-07154, and PHY 02-16873:  Physics Frontier Center / Joint Institute for
Nuclear Astrophysics (JINA), both awarded by the National Science Foundation.
	C.A.P.\ acknowledges partial funding for this work from NASA grants NAG 
5-13057 and NAG 5-13147.




\appendix
\section{DATA TABLES}

	Tables \ref{tab:dat1} and \ref{tab:dat2} contain the photometric and
spectroscopic measurements for the 2414 2MASS-selected BHB candidates.  Our survey
boundaries include 100 objects previously observed as part of Papers I and II.  We
include the previously published objects in Tables \ref{tab:dat1} and \ref{tab:dat2}
for completeness.  Tables \ref{tab:dat1} and \ref{tab:dat2} are presented in their
entirety in the electronic edition of the Astronomical Journal.  A portion of the
tables are shown here for guidance regarding their format and content.

	Table \ref{tab:dat1} summarizes photometry and positions.  Column (1) is our
identifier.  The designation CHSS stands for Century Halo Star Survey and is chosen
to be unique from previous surveys.  Column (2) is the J2000 right ascension in
hours, minutes, and seconds.  Column (3) is the J2000 declination in degrees,
arcminutes, and arcseconds.  Column (4) is the extinction-corrected 2MASS $J_0$
magnitude.  Columns (5) and (6) are the extinction-corrected 2MASS colors $(J-H)_0$
and $(H-K)_0$.  Column (7) is the $E(B-V)$ reddening value from \citet{schlegel98}.
Columns (8) and (9) are the Galactic coordinates, in degrees.  Column (10) is our
$BV0$ color predicted from 2MASS photometry and Balmer line strengths (and SDSS
photometry, where available).

	Table \ref{tab:dat2} summarizes the spectroscopic and stellar parameters.  
We include all the DA white dwarfs and subdwarfs in this table, but we omit their
stellar parameters as our analysis is meaningless for these objects.  We also omit
stellar parameters for a few dozen objects with unusually low signal-to-noise
spectra.  Column (1) is our identifier.  Column (2) is the heliocentric radial
velocity in km s$^{-1}$.  Column (3) is the spectral type, where B0=10, A0=20,
F0=30, and so forth.  Column (4) is the effective temperature in K.  Column (5) is
the surface gravity in cm s$^{-2}$.  Column (6) is the metallicity given as the
logarithmic [Fe/H] ratio relative to the Sun.  Column (7) is our classification:  
BHB = blue horizontal branch star, BHB/A = possible blue horizontal branch star with
[Fe/H] $>-0.6$, A = high surface gravity, early A-type star, DA = DA white dwarf, sd
= subdwarf. Column (8) is the absolute $M_V$ magnitude estimated from Equations
\ref{eqn:mvbhb} and \ref{eqn:mvbs} for BHB and A stars, respectively.  Column (9) is
the estimated distance in kpc.  Absolute magnitude and distance estimates are only 
provided for BHB and A-type stars, as described in Section 3.5.


\begin{deluxetable}{lrrccccccc}		
\tablecolumns{10}	\tablewidth{0pt}
\tabletypesize{\small}
\tablecaption{PHOTOMETRY\label{tab:dat1}}
\tablehead{
	\colhead{} & \colhead{} & \colhead{} & 	\colhead{$J_0$} & 
	\colhead{$(J-H)_0$} & \colhead{$(H-K)_0$} & \colhead{$E(\bv)$} & 
	\colhead{$l$} & \colhead{$b$} & \colhead{$BV0$} \\
        \colhead{ID} & \colhead{$\alpha_{\rm J2000}$} & \colhead{$\delta_{\rm J2000}$} &
        \colhead{(mag)} & \colhead{(mag)} & \colhead{(mag)} & \colhead{(mag)} &
        \colhead{(deg)} & \colhead{(deg)} & \colhead{(mag)} \\
        \colhead{(1)} & \colhead{(2)} & \colhead{(3)} &
        \colhead{(4)} & \colhead{(5)} & \colhead{(6)} & \colhead{(7)} & 
	\colhead{(8)} & \colhead{(9)} & \colhead{(10)} \\
}
	\startdata
CHSS 3014 &  0:01:29.2 & 16:01:51 & $13.26 \pm 0.026$ & $ 0.06 \pm 0.04$ & $ 0.07 \pm 0.05$ & 0.037 & 105.787 & $-45.172$ & $ 0.17$ \\
CHSS 3015 &  0:01:31.4 & 18:36:09 & $14.20 \pm 0.028$ & $ 0.04 \pm 0.05$ & $ 0.08 \pm 0.07$ & 0.033 & 106.753 & $-42.696$ & $ 0.24$ \\
CHSS 3016 &  0:01:32.9 & 22:58:26 & $15.01 \pm 0.040$ & $-0.18 \pm 0.11$ & $ 0.02 \pm 0.17$ & 0.079 & 108.220 & $-38.468$ & $ 0.08$ \\
CHSS 3017 &  0:01:59.2 & 25:01:07 & $13.77 \pm 0.027$ & $-0.05 \pm 0.05$ & $ 0.01 \pm 0.07$ & 0.088 & 108.964 & $-36.508$ & $ 0.12$ \\
	\enddata 
  \tablecomments{Table \ref{tab:dat1} is presented in its entirety
in the electronic edition of the Astronomical Journal.  A portion is shown
here for guidance and content.}
 \end{deluxetable}

\begin{deluxetable}{ccccccccc}	
\tabletypesize \small
\tablecaption{SPECTROSCOPIC AND STELLAR PARAMETERS\label{tab:dat2}}
\tablecolumns{9}
\tablehead{
	\colhead{} & \colhead{$v_{radial}$} & \colhead{} & 
	\colhead{$T_{eff}$} & \colhead{$\log{g}$} & \colhead{} &
	\colhead{} & \colhead{$M_V$} & \colhead{Dist} \\
	\colhead{ID} & \colhead{(km s$^{-1}$)} & \colhead{Type} &
	\colhead{(K)} & \colhead{(cm s$^{-2}$)} & \colhead{[Fe/H]} &
	\colhead{Class} & \colhead{(mag)} & \colhead{(kpc)} \\
	\colhead{(1)} & \colhead{(2)} & \colhead{(3)} &
	\colhead{(4)} & \colhead{(5)} & \colhead{(6)} & 
	\colhead{(7)} & \colhead{(8)} & \colhead{(9)} \\
}
	\startdata
CHSS 3014 & $-163 \pm 11$ & $20.3 \pm 2.1$ &    7644 &    3.26 & $-1.02$ &    BHB & $0.61 \pm 0.08$ & $ 4.01 \pm 0.31$ \\
CHSS 3015 & $ -54 \pm 12$ & $27.1 \pm 1.9$ &    7356 &    3.54 & $-0.65$ &\nodata &       \nodata   &        \nodata   \\
CHSS 3016 & $-247 \pm 13$ & $20.7 \pm 2.4$ &    8388 &    3.51 & $-1.34$ &    BHB & $0.70 \pm 0.10$ & $ 7.85 \pm 0.64$ \\
CHSS 3017 & $ -10 \pm 14$ & $22.2 \pm 1.7$ &    8195 &    4.20 & $-0.18$ &      A & $1.90 \pm 0.20$ & $ 2.67 \pm 0.32$ \\
	\enddata	
  \tablecomments{Table \ref{tab:dat2} is presented in its entirety
in the electronic edition of the Astronomical Journal.  A portion is shown
here for guidance and content.}
 \end{deluxetable}

\end{document}